\documentclass[a4paper,preprintnumbers,showpacs,twocolumn,superscriptaddress,nofootinbib,amsmath,amssymb]{revtex4-1}
\usepackage{color,hyperref}
\usepackage{times}
\usepackage{mathrsfs}
\hypersetup{colorlinks=true,linkcolor=blue,citecolor=magenta,filecolor=magenta,urlcolor=blue}
\usepackage{graphicx}
\usepackage{dcolumn}
\usepackage{bm}
\usepackage{orcidlink}

\newcommand{\rh}{r_{\rm h}}
\newcommand{\bc}{b_{\rm c}}
\newcommand{\bg}{b_{\rm g}}
\newcommand{\Qmax}{Q_{\rm max}}
\newcommand{\dsdo}{\frac{d\sigma}{d\Omega_{\rm D-2}}}

\begin{document}

\title{Scattering of massless scalar waves by higher-dimensional regular black holes in general relativity coupled to nonlinear electrodynamics}

\author{Bobir Toshmatov\,\orcidlink{0000-0002-7853-186X}}
\email{toshmatov@astrin.uz}
\affiliation{New Uzbekistan University, Movarounnahr street 1, Tashkent 100000, Uzbekistan}
\affiliation{Ulugh Beg Astronomical Institute, Astronomy street 33, Tashkent 100052, Uzbekistan}

\author{Askar Abdikamalov\,\orcidlink{0000-0002-7671-6457}}
\email{a.abdikamalov@newuu.uz}
\affiliation{New Uzbekistan University, Movarounnahr street 1, Tashkent 100000, Uzbekistan}
\affiliation{Ulugh Beg Astronomical Institute, Astronomy street 33, Tashkent 100052, Uzbekistan}

\author{Sardor Tojiev\,\orcidlink{0000-0002-5660-7800}}
\email{sardor@astrin.uz}
\affiliation{Tashkent State University of Economics, Islom Karimov street 49, 100066, Uzbekistan}
\affiliation{Institute of Fundamental and Applied Research, National Research University TIIAME, Kori Niyoziy 39, Tashkent 100000, Uzbekistan}

\begin{abstract}
We study the scattering of planar massless scalar waves by the family of higher-dimensional regular black holes in general relativity coupled to nonlinear electrodynamics, which unifies the Bardeen and Hayward models in any dimension. The scalar field is a neutral test field that propagates in the bulk, and in the geometric-optics limit, its rays are the null geodesics of the spacetime, so that the unstable circular null geodesics of the background geometry control the scattering. We combine the classical scattering of null geodesics, the glory approximation, which we generalize to $D$ dimensions, and the partial-wave analysis in $D=5,6,7$. We show that at a fixed frequency, the wave cross section diverges in the forward direction as $\theta^{-4}$ in any dimension, with a coefficient fixed by the mass, while the steeper classical law $\theta^{-(D-2)^{2}/(D-3)}$ appears only at larger angles and higher frequencies. The charge parameter decreases the critical and the glory impact parameters and widens the glory fringes, and in higher dimensions, the glory pattern has fewer and wider fringes and a weaker backward intensity. For the matched configurations studied here, backward intensity and low-frequency absorption provide the clearest differences, while the intermediate-to-high-frequency spectra are close. 
\end{abstract}

\maketitle

\section{Introduction}\label{sec:intro}

The scattering of waves by black holes (BHs) has a long history~\cite{Matzner1968,Vishveshwara1970,Sanchez1978,Futterman1988}. The scattering of the scalar, electromagnetic (EM), spinor, and gravitational fields has been studied in the Schwarzschild, Reissner-Nordstr\"om (RN), and Kerr spacetimes~\cite{Crispino2009RN,Crispino2009EM,Dolan2008,Macedo2013,Crispino2014,deOliveira:2017gxs}, and it has been understood that the main features of the scattering cross section can be explained by the properties of null geodesics. The cross section diverges in the forward direction because of the long-range character of the gravitational field, and it shows interference fringes in the backward direction, which are known as the glory~\cite{Matzner1985,Futterman1988}. In the eikonal regime, the glory and the quasinormal modes of the scalar perturbations are both related to the unstable circular null geodesics of the spacetime.

General relativity predicts the existence of curvature singularities inside BHs~\cite{Penrose1965,Hawking1973}, and regular BH models have been proposed as a way to avoid them. The first regular BH was suggested by Bardeen~\cite{Bardeen1968}, and later it was shown that such solutions can be obtained in general relativity coupled to nonlinear electrodynamics (NED)~\cite{AyonBeato1998,AyonBeato2000,PhysRevLett.85.4641,Bronnikov2001,Dymnikova2004,Hayward2006,FanWang2016,Toshmatov2017,Toshmatov2018comment,Diez:2026bmp}. In these models, the central singularity is replaced by a de Sitter core, and the charge parameter plays the role of a length scale that regularizes the geometry. The quasinormal modes, the relaxation of perturbations, and the motion of particles and light in the regular spacetimes have been studied in detail~\cite{Flachi2013,Toshmatov2015,Toshmatov2021light,Toshmatov2026AoP}. The scattering of massless scalar waves by the Bardeen BH was investigated in Refs.~\cite{Macedo2014,Macedo2015,Macedo2016}, where it was shown that for some values of the charge the scattering pattern is very similar to that of the RN BH, and that the glory is controlled by a single geometric quantity, the glory impact parameter. This was later extended to other regular BHs and to massive fields~\cite{Paula2020,Paula2022,LimaJunior2022,Paula2026,Tang2026}. Therefore, the question of whether the wave scattering can distinguish the exterior geometry of a regular BH from that of a singular one has a substantial literature in four dimensions.

One point should be stressed here. In the NED, the scalar and gravitational perturbations propagate on the spacetime geometry, and in the geometric-optics limit, their rays are the null geodesics of the spacetime. The light rays and the EM perturbations, instead, follow the null geodesics of an effective (optical) geometry, which is determined by the NED Lagrangian~\cite{Novello2000,Obukhov:2002xa,Chaverra:2016ttw,Moreno:2002gg,Toshmatov2018axial,Toshmatov2018polar,Toshmatov2019relax}. As a result, the photonsphere, which is the unstable circular orbit of the light rays, does not coincide with the unstable circular null geodesic of the spacetime. In this paper, we consider a neutral, minimally coupled, massless test scalar field that propagates in the bulk, and therefore, the relevant quantity is the unstable circular null geodesic of the background geometry, not the photonsphere, as we have shown in our papers on the EM perturbations and their relaxations~\cite{Toshmatov2018axial,Toshmatov2018polar,Toshmatov2019relax}. The scattering of the EM waves, in which the effective geometry must be taken into account, is a separate problem, and it will be studied in a forthcoming paper.

Extra spatial dimensions are motivated by the gap between gravity and the other interactions~\cite{ArkaniHamed1998,Antoniadis1998,Randall1999,Emparan2008}, and in such scenarios, small BHs can be described by higher-dimensional solutions~\cite{Tangherlini1963,Emparan2008}. Wave scattering by higher-dimensional BHs has been studied in this context~\cite{Harris2003,Kanti2004,Decanini2011a,Decanini2011b,Crispino2010HD,Ahmedov2021}. The dimension of spacetime affects the scattering in two ways. It changes the long-range behavior of the potential, which becomes $\propto r^{-(D-3)}$ and therefore much shorter ranged, and it changes the geometry of the wave front, since the planar wave is expanded over the harmonics on $S^{D-2}$. As a result, the partial-wave series behaves differently from the four-dimensional one, and the relation between the wave and the classical descriptions of the forward scattering is not the same as in four dimensions.

Recently, a family of electrically charged, higher-dimensional regular BHs in general relativity coupled to the NED has been obtained~\cite{Toshmatov2026HD,Toshmatov:2026xtc}, with the metric function
\begin{equation}\label{eq:f0}
f(r)=1-\frac{M r^{2}}{(r^{\gamma}+Q^{\gamma})^{(D-1)/\gamma}},
\end{equation}
which unifies the Bardeen ($\gamma=2$) and Hayward ($\gamma=3$) models in any dimension $D\geq4$. In this paper, we study how these BHs scatter planar massless scalar waves and whether their exterior geometry can be distinguished from that of the Reissner-Nordstr\"om-Tangherlini (RNT) BH. We follow the approach of Refs.~\cite{Macedo2014,Macedo2015,Macedo2016,Paula2020,Paula2026} for the four-dimensional regular BHs, extended to the bulk propagation in $D$ dimensions, and we combine the classical scattering of null geodesics, the semiclassical glory approximation, and the numerical partial-wave analysis. We focus on the Maxwellian cases, $D=5$ with $\gamma=2$ and $D=6$ with $\gamma=3$, and we also vary $\gamma$, the frequency, and the dimension up to $D=7$ to understand the role of each parameter.

The paper is organized as follows. In Sec.~\ref{sec:bh}, we review the spacetime and its basic properties. In Sec.~\ref{sec:geo}, we study the null geodesics, the deflection angle, the glory, and the classical scattering. In Sec.~\ref{sec:wave}, we introduce the scalar field, the partial-wave expansion, and the semiclassical approximations. In Sec.~\ref{sec:results}, we discuss the results, and in Sec.~\ref{sec:conclusion}, we summarize them. We use the signature $(-,+,\dots,+)$ and the units $G_{D}=1=c$. Lengths and frequencies are measured in units of $M^{1/(D-3)}$ and $M^{-1/(D-3)}$, respectively, and the differential cross sections in units of $M^{(D-2)/(D-3)}$.

\section{Higher-dimensional regular black holes}\label{sec:bh}

The $D$-dimensional static, spherically symmetric spacetime is described by
\begin{equation}\label{eq:metric}
ds^{2}=-f(r)dt^{2}+\frac{dr^{2}}{f(r)}+r^{2}d\Omega_{D-2}^{2},
\end{equation}
where $f(r)$ is given by Eq.~(\ref{eq:f0}), $M$ is the mass parameter, $Q$ is the charge parameter, $\gamma>0$ characterizes the model \cite{Toshmatov2026HD}, and $d\Omega_{D-2}^{2}$ is the line element of the unit $(D-2)$-sphere. The mass parameter is related to the ADM mass by 
\begin{eqnarray}
    M_{\rm ADM}=\frac{(D-2)\Omega_{D-2}}{16\pi}M\ ,
\end{eqnarray}
where $\Omega_{D-2}$ is the area of the unit sphere. As it was presented in \cite{Toshmatov2026HD} that at large distances, the metric function behaves as 
\begin{equation}\label{eq:fasymp}
f\simeq1-\frac{M}{r^{D-3}}+\frac{(D-1)MQ^{\gamma}}{\gamma\,r^{D+\gamma-3}},
\end{equation}
i.e., as the Schwarzschild-Tangherlini one with a charge correction, while near the center, $r\ll Q$, the spacetime is de Sitter-like, 
\begin{eqnarray}
    f\simeq1-\frac{2\Lambda_{\rm eff}}{(D-1)(D-2)}r^{2}\ ,
\end{eqnarray}
with 
\begin{eqnarray}
    \Lambda_{\rm eff}=\frac{(D-1)(D-2)M}{2Q^{D-1}}\ ,
\end{eqnarray}
and the geometry is regular. The source is an electric field of the NED with the Lagrangian density $\mathcal{L}(F)$, $F=F_{\mu\nu}F^{\mu\nu}$, and the Einstein equations determine $\mathcal{L}$ and $\mathcal{L}_{F}$ in terms of the metric function~\cite{Toshmatov2026HD}. For the family (\ref{eq:f0}), $\mathcal{L}_{F}$ tends to a constant at infinity, i.e., the NED has the Maxwellian weak-field limit, only for $\gamma=D-3$. In this case, the charge correction in Eq.~(\ref{eq:fasymp}) decays in the same way as in the RNT spacetime,
\begin{equation}\label{eq:RNT}
f_{\rm RNT}=1-\frac{M}{r^{D-3}}+\frac{q^{2}}{r^{2(D-3)}},
\end{equation}
and the comparison of the leading terms gives 
\begin{eqnarray}
    q^{2}=\frac{D-1}{D-3}MQ^{D-3}
\end{eqnarray}
which relates the charge parameter $Q$ of the regular BH to the charge parameter $q$ of the RNT metric. The Maxwellian regular black hole cases are $\gamma=2$ for $D=5$ (Bardeen-like) and $\gamma=3$ for $D=6$ (Hayward-like), and they are the main subject of this paper.

As it was shown in \cite{PhysRevLett.85.4641,Bronnikov2001} that for the electrically charged regular BHs, the Lagrangian density necessarily consists of two branches
joined at a cusp that makes it double valued. Therefore,  the theory we presented is defined through the Hamiltonian density with fixed couplings~\cite{Toshmatov2026HD,Toshmatov:2026xtc}. These outer and inner branches of the Lagrangian coincide at the radius $r_{\rm b}=Q[(\gamma+D-2)/(\gamma+1)]^{1/\gamma}$, which in four dimensions is always inside the horizon, but for $D\geq5$ lies outside the horizon of the near-extremal BHs. Whether the backgrounds in this range are acceptable solutions of the NED with a well-posed dynamics is an open question. In the present paper, the scalar field is a neutral test field on the prescribed smooth exterior metric, and its wave equation is well defined for all $\alpha\leq1$ regardless of this question. Our results therefore concern the exterior geometry only, and the scan over the charge at a fixed mass parameter is a comparison across the family of regular metrics rather than within one theory, since the coupling of the Hamiltonian density changes with $Q$ at fixed $M$~\cite{Toshmatov:2026xtc}.

The spacetime represents the BH for $Q\leq\Qmax$ and no horizon object for $Q>\Qmax$, where~\cite{Toshmatov2026HD}
\begin{equation}\label{eq:Qmax}
\Qmax=(D-3)^{1/\gamma}\left[\frac{4}{(D-1)^{D-1}}\right]^{1/[\gamma(D-3)]}M^{1/(D-3)}.
\end{equation}
The horizon radius $\rh$ is the largest root of $f(\rh)=0$. Since we perform calculations for different spacetimes in various spacetime dimensions, the maximum charge, $Q_{\rm max}$, varies with these parameters. Therefore, to facilitate comparisons across different spacetimes and dimensions, we adopt the normalized charge.
\begin{equation}\label{eq:alpha}
\alpha\equiv Q/\Qmax\,,\qquad 0\leq\alpha\leq1,
\end{equation}
where $\alpha=0$ corresponds to the Schwarzschild-Tangherlini BH and $\alpha=1$ to the extremal one, whose Hawking temperature vanishes \cite{Toshmatov:2026xtc}. For the RNT BH, $\alpha=q/q_{\max}$ with $q_{\max}=M/2$. For the comparison with the singular RNT BH, we mainly compare the configurations with the same $\alpha$, since $\alpha$ measures how close a BH is to extremality. For the Maxwellian models, we also compare the configurations with the same charge $q$.

\section{Null geodesics and classical scattering}\label{sec:geo}

\subsection{Circular null geodesics and deflection angle}

\begin{figure}[!htb]
\includegraphics[width=\columnwidth]{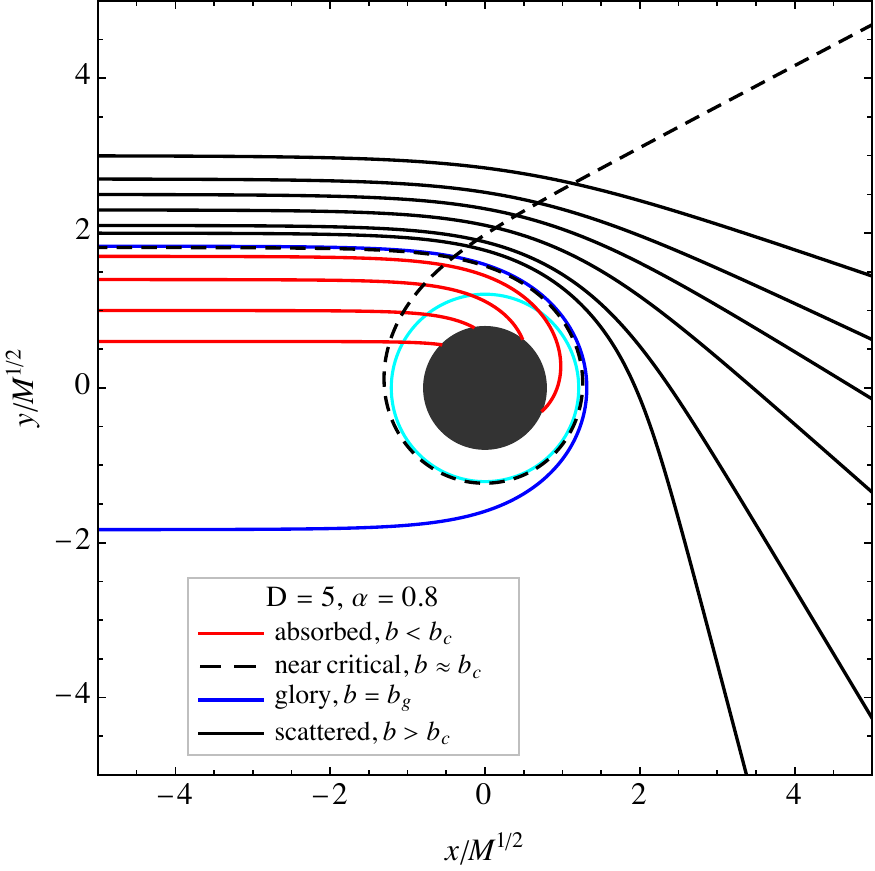}
\caption{Null geodesics around the five-dimensional Bardeen-like regular BH with $\alpha=0.8$, namely, absorbed rays (red), the glory ray (blue), an almost critical ray (dashed), and scattered rays (black). The black disk is the event horizon, and the cyan circle is the unstable circular null geodesic.}
\label{fig:traj}
\end{figure}

As mentioned in previous section, the scalar waves propagate along the null geodesics of the spacetime (\ref{eq:metric}). Null geodesics in the equatorial plane are determined by the orbit equation that reads as following in terms of $u=1/r$ \cite{Misner:1973prb}:
\begin{equation}\label{eq:radial}
\left(\frac{du}{d\varphi}\right)^{2}=\frac{1}{b^{2}}-f(1/u)u^{2}\ ,
\end{equation}
where $b=L/E$ is the impact parameter, with $E$ and $L$ being the energy and angular momentum of the massless particle, respectively. The unstable circular null geodesic is the outermost root $r_{c}$ of $2f(r_{c})=r_{c}f'(r_{c})$, which for the metric (\ref{eq:f0}) gives
\begin{equation}\label{eq:rc}
(D-1)M\,r_{c}^{\gamma+2}=2\left(r_{c}^{\gamma}+Q^{\gamma}\right)^{(D+\gamma-1)/\gamma},
\end{equation}
and the critical impact parameter is $\bc=r_{c}/\sqrt{f(r_{c})}$. Rays with $b<\bc$ are absorbed by the BH, and rays with $b>\bc$ are scattered, as shown in Fig.~\ref{fig:traj}. The circular orbit is unstable, and the deviation of a nearby ray grows as $e^{\lambda_{c}t}$ with the Lyapunov exponent
\begin{equation}\label{eq:lyap}
\lambda_{c}=\sqrt{\frac{f(r_{c})\left[2f(r_{c})-r_{c}^{2}f''(r_{c})\right]}{2r_{c}^{2}}}\,,\qquad
\eta_{c}\equiv\lambda_{c}\bc\,,
\end{equation}
where the dimensionless combination $\eta_{c}$ controls the strong-deflection behavior~\cite{Cardoso2009}. For the Tangherlini BH, $\eta_{c}=\sqrt{D-3}$, and the charge decreases $\eta_{c}$ (Tab.~\ref{tab:geo}).

If we assume that the source of the scalar wave and receiver are far away from the black hole, the deflection angle of a scattered ray can be found from
\begin{equation}\label{eq:Theta}
\Theta(b)=2\int_{0}^{u_{0}}\frac{du}{\sqrt{b^{-2}-f(1/u)u^{2}}}-\pi,
\end{equation}
where $u_0$ is the turning point. At large impact parameters, i.e., in the weak gravitational field regime, the first-order contributions of the mass term and of the charge correction of Eq.~(\ref{eq:fasymp}) are
\begin{equation}\label{eq:weak}
\Theta\simeq\frac{\sqrt{\pi}\,\Gamma(\tfrac{D}{2})}{\Gamma(\tfrac{D-1}{2})}\frac{M}{b^{D-3}}-\frac{\sqrt{\pi}\,\Gamma(\tfrac{D+\gamma}{2})}{\Gamma(\tfrac{D+\gamma-1}{2})}\frac{(D-1)MQ^{\gamma}}{\gamma\,b^{D+\gamma-3}}\ .
\end{equation}
This shows that the regular BH deflects the rays less than the Schwarzschild-Tangherlini BH of the same mass. For $\gamma=D-3$, the second-order mass term appears at the same order as the charge term, and the complete expansion through this order is given in Appendix~\ref{app:series}. When $b$ approaches $\bc$ from above, i.e., in the strong gravitational field regime, the deflection angle diverges logarithmically~\cite{Bozza:2001xd,Bozza:2002zj},
\begin{equation}\label{eq:strong}
\Theta(b)\simeq-\frac{1}{\eta_{c}}\ln\left(\frac{b}{\bc}-1\right)+\bar{b},
\end{equation}
where $\bar b$ is a constant, which is the signature of the rays that wind around the unstable circular null geodesic before escaping.

The ray with $\Theta=\pi$ is scattered exactly backwards, and its impact parameter $\bg$, the glory impact parameter, is determined by $\Theta(\bg)=\pi$. The rays deflected by $(2k+1)\pi$ also go backwards, and their impact parameters approach $\bc$ exponentially, $b_{g}^{(k)}/\bc-1\propto e^{-(2k+1)\pi\eta_{c}}$. Their contribution to the backward scattering is proportional to $|db/d\theta|_{\theta=\pi}=\eta_{c}\bc(b/\bc-1)$, so that the $k$-th ray is suppressed by $e^{-2\pi k\eta_{c}}$ with respect to the glory ray. This factor is very small for the Tangherlini BHs, but it grows as the charge decreases $\eta_{c}$, and for the extremal five-dimensional BH, the amplitude of the first winding ray is a few percent of that of the glory ray.

\begin{figure*}[!htb]
\includegraphics[width=0.48\linewidth]{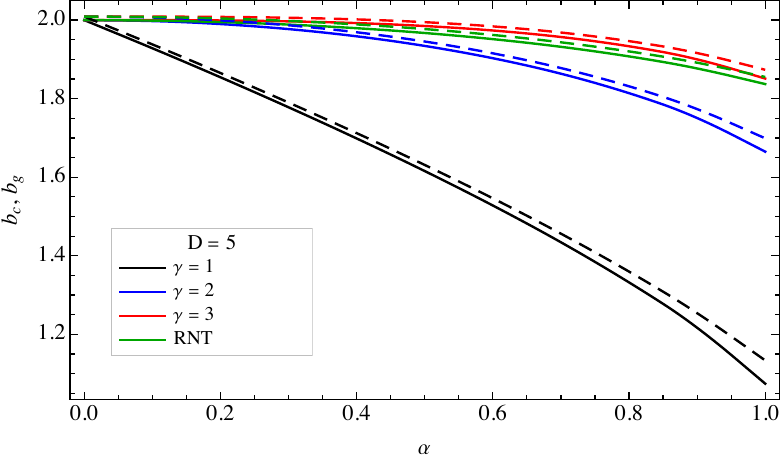}
\includegraphics[width=0.48\linewidth]{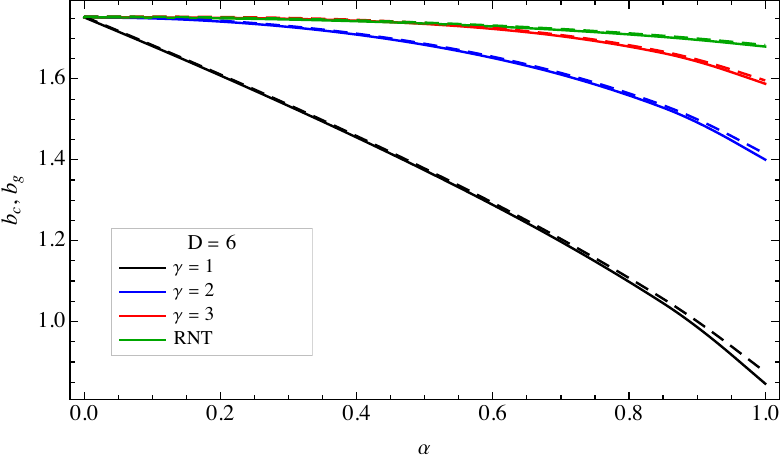}
\caption{Critical impact parameter $\bc$ (solid) and glory impact parameter $\bg$ (dashed) as functions of the normalized charge for $D=5$ and $D=6$ (left and right panels), $\gamma=1$ (black), $\gamma=2$ (blue), $\gamma=3$ (red), and the RNT BH (green).}
\label{fig:bc}
\end{figure*}

\begin{figure*}[!htb]
\includegraphics[width=0.48\linewidth]{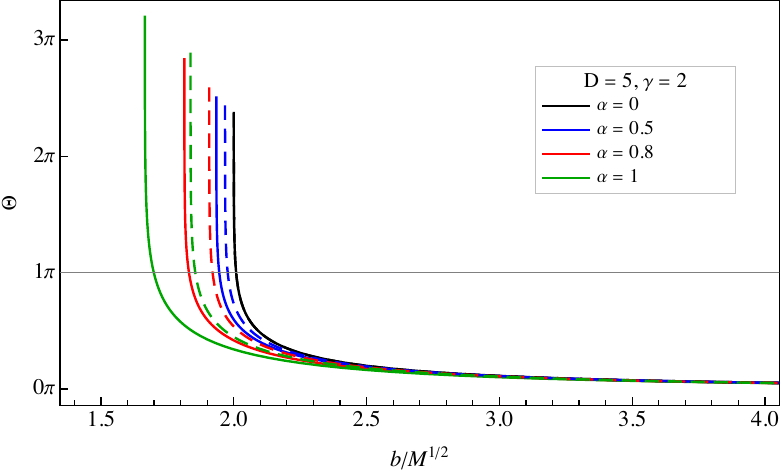}
\includegraphics[width=0.48\linewidth]{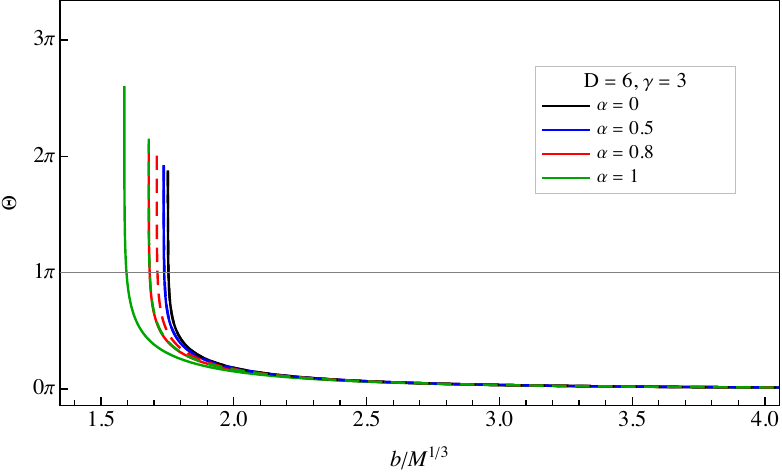}
\caption{Deflection angle of null geodesics for $D=5$, $\gamma=2$ and $D=6$, $\gamma=3$ (left and right panels). The dashed curves correspond to the RNT BH with the same $\alpha$, and the horizontal gray line marks $\Theta=\pi$, which determines $\bg$.}
\label{fig:defl}
\end{figure*}

Fig.~\ref{fig:bc} shows $\bc$ and $\bg$ as functions of $\alpha$, and Tab.~\ref{tab:geo} lists some of their values together with $\eta_{c}$. Both impact parameters decrease with the charge parameter, since the charge screens the gravitational attraction, and $\bg$ is slightly larger than $\bc$. At the same normalized charge, the regular BHs with $\gamma\leq D-3$ have smaller $\bc$ and $\bg$ than the RNT BH, while at the same charge $q$, they are slightly larger, as in four dimensions~\cite{Macedo2014,Macedo2016}. The critical impact parameter is smaller in higher dimensions and for smaller $\gamma$, since a smaller $\gamma$ means a slower decay of the effective matter distribution and a stronger screening of the mass at the circular null geodesic. In Fig.~\ref{fig:defl}, we present the deflection angle of the scalar wave as a function of the impact parameter. It shows that the deflection angle diverges at $b=\bc$, decreases as $b^{-(D-3)}$ at large $b$, and it decreases with $\alpha$ for a fixed $b$. The intersections of the curves with the line $\Theta=\pi$ give $\bg$, and the strong-deflection region moves closer to the BH as the charge increases.

\begin{table}[!htb]
\caption{Event horizon radius, critical impact parameter, glory impact parameter (in units of $M^{1/(D-3)}$), and the parameter $\eta_{c}$ of the unstable circular null geodesic. The last two rows are the Maxwellian regular BHs with the same charge as the extremal RNT BH, $q=M/2$.}
\label{tab:geo}
\begin{ruledtabular}
\begin{tabular}{ccccccc}
$D$ & Model & $\alpha$ & $\rh$ & $\bc$ & $\bg$ & $\eta_{c}$ \\
\hline
5 & $\gamma=2$ & 0.5 & 0.9330 & 1.9343 & 1.9454 & 1.341 \\
5 & $\gamma=2$ & 1 & 0.5000 & 1.6651 & 1.6990 & 0.924 \\
5 & RNT & 0.5 & 0.9659 & 1.9671 & 1.9772 & 1.377 \\
5 & RNT & 1 & 0.7071 & 1.8371 & 1.8553 & 1.155 \\
6 & $\gamma=3$ & 0.5 & 0.9765 & 1.7359 & 1.7382 & 1.693 \\
6 & $\gamma=3$ & 1 & 0.6011 & 1.5880 & 1.5962 & 1.242 \\
6 & RNT & 0.5 & 0.9771 & 1.7369 & 1.7392 & 1.694 \\
6 & RNT & 1 & 0.7937 & 1.6799 & 1.6838 & 1.500 \\
\hline
5 & $\gamma=2$ & 0.707 ($q=M/2$) & 0.8536 & 1.8604 & 1.8750 & 1.251 \\
6 & $\gamma=3$ & 0.772 ($q=M/2$) & 0.9018 & 1.6879 & 1.6912 & 1.569 \\
\end{tabular}
\end{ruledtabular}
\end{table}

\subsection{Classical scattering cross section}

\begin{figure*}[!htb]
\includegraphics[width=0.48\linewidth]{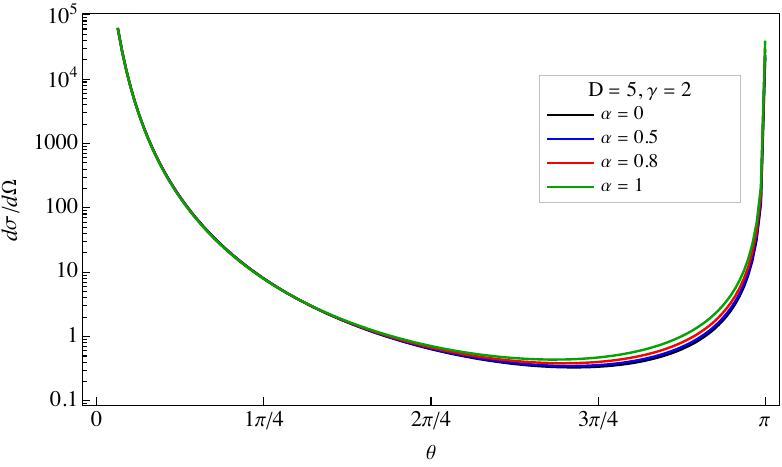}
\includegraphics[width=0.48\linewidth]{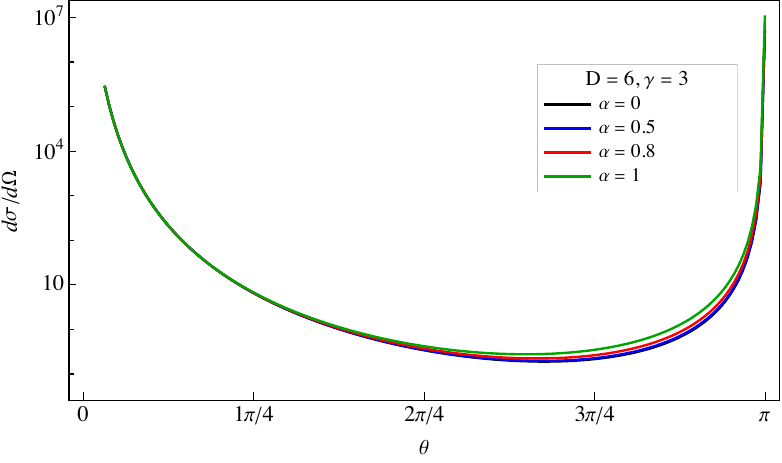}
\caption{Classical scattering cross section (\ref{eq:classical}) for $D=5$, $\gamma=2$ and $D=6$, $\gamma=3$ (left and right panels).}
\label{fig:class}
\end{figure*}

Now let us consider a uniform flux of null geodesics that is scattered into the solid angle $d\Omega_{D-2}$. The rays that arrive at the angle $\theta$ are those with $\Theta(b)=\theta+2\pi k$ and those with $\Theta(b)=2\pi(k+1)-\theta$, which pass on the other side of the BH, and the conservation of the flux gives~\cite{LimaJunior2022}
\begin{equation}\label{eq:classical}
\dsdo=\frac{1}{\sin^{D-3}\theta}\sum_{j}b_{j}^{D-3}\left|\frac{db_{j}}{d\theta}\right|,
\end{equation}
where $b_{j}$ runs over all the impact parameters deflected into $\theta$. The branch $\Theta=\theta$ dominates at small and intermediate angles, while the branches $\Theta=\theta$ and $\Theta=2\pi-\theta$ become equally important near the backward direction, since both approach the glory ray. At small angles, the weak-field expression (\ref{eq:weak}) gives $b\propto\theta^{-1/(D-3)}$, so that
\begin{equation}\label{eq:small}
\dsdo\propto\theta^{-(D-2)^{2}/(D-3)},
\end{equation}
independently of the charge, with a coefficient fixed by the mass. The exponent grows with the dimension, and in $D=4$, the Rutherford-like law $\theta^{-4}$ is recovered. Eq. (\ref{eq:small}) is the geometric-optics law, and we show in Sec.~\ref{sec:wave} that it is not the forward asymptotics of the wave cross section at a fixed frequency. The classical cross section is shown in Fig.~\ref{fig:class}. At larger angles, it grows with $\alpha$, since the deflection function is flatter for smaller $\eta_{c}$ and the Jacobian $|db/d\theta|$ is larger, and it diverges at $\theta=\pi$, where $\sin\theta\to0$. This divergence is regularized by the wave nature of the field.

\section{Massless scalar field}\label{sec:wave}

\subsection{Wave equation and effective potential}

\begin{figure*}[!htb]
\includegraphics[width=0.48\linewidth]{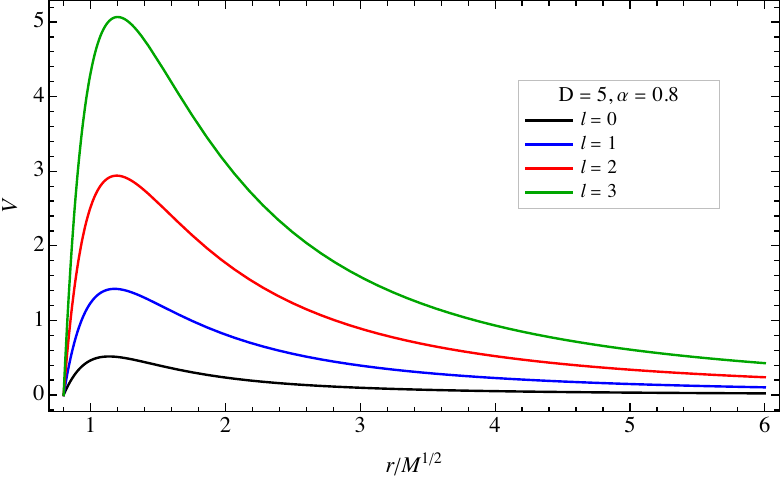}
\includegraphics[width=0.48\linewidth]{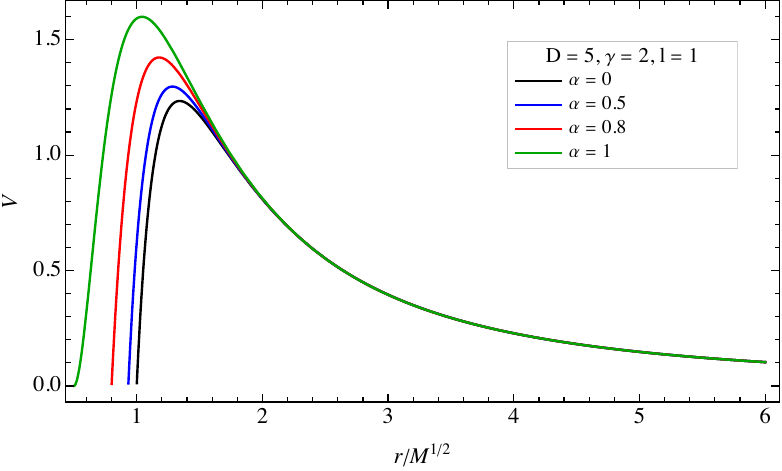}
\caption{Effective potential of the scalar field for the five-dimensional Bardeen-like regular BH.}
\label{fig:V}
\end{figure*}

The neutral, minimally coupled, massless scalar field in the curved spacetime is governed by the Klein-Gordon equation $\Box\Phi=0$, where the wave function $\Phi$ is the function of coordinates of the spacetime. To separate the variables in the Klein-Gordon equation, we adopt the hyperspherical harmonics on $S^{D-2}$ and choose the wave function as 
\begin{eqnarray}
    \Phi=\sum_{l}r^{-(D-2)/2}\psi_{l}(r)Y_{l}(\theta_{i})e^{-i\omega t}\ .
\end{eqnarray}
Thus, we obtain
\begin{equation}\label{eq:RW}
\frac{d^{2}\psi_{l}}{dr_{*}^{2}}+\left[\omega^{2}-V(r)\right]\psi_{l}=0,\qquad dr_{*}=\frac{dr}{f},
\end{equation}
with the effective potential
\begin{equation}\label{eq:V}
V=f\left[\frac{l(l+D-3)}{r^{2}}+\frac{(D-2)(D-4)}{4r^{2}}f+\frac{(D-2)f'}{2r}\right].
\end{equation}
At large distances, the potential behaves as $V\to(\nu^{2}-1/4)/r^{2}$ with $\nu=l+(D-3)/2$, and it vanishes at the horizon. For large $l$, the maximum of the barrier is located near the unstable circular null geodesic, $V_{\max}\simeq\nu^{2}/\bc^{2}$, which is the wave counterpart of the critical impact parameter. A wave with $\nu<\omega\bc$ passes over the barrier and is mostly absorbed, while the waves with $\nu>\omega\bc$ are mostly scattered. The behavior of the effective potential is shown in Fig.~\ref{fig:V}. It shows that the height of a barrier grows with $l$, and it becomes higher and moves toward the horizon as the charge increases. The effective potential for all $\alpha$, $l$, and $D$ coincide at large $r$, where the mass term dominates, which is the origin of the universality of the high-$l$ phase shifts discussed below.

\subsection{Scattering amplitude}

To solve the Klein-Gordon equation, one must impose the boundary condition. We choose the boundary condition such that a planar wave comes from infinity and is purely ingoing at the horizon,
\begin{equation}\label{eq:bc_wave}
\psi_{l}\to
\begin{cases}
e^{-i\omega r_{*}}, & r_{*}\to-\infty,\\[1mm]
A_{l}^{-}e^{-i\omega r_{*}}+A_{l}^{+}e^{+i\omega r_{*}}, & r_{*}\to+\infty,
\end{cases}
\end{equation}
which defines the scattering matrix $S_{l}=(-1)^{l}i^{D-2}A_{l}^{+}/A_{l}^{-}$ and the complex phase shifts $\delta_{l}=(2i)^{-1}\ln S_{l}$, with ${\rm Im}\,\delta_{l}\geq0$ because of the absorption by the horizon. In $D$ dimensions, the plane wave is expanded in the Gegenbauer polynomials $C_{l}^{\lambda}$ with $\lambda=(D-3)/2$, and the scattering amplitude and the differential cross section are
\begin{align}\label{eq:dsdo}
h(\theta)&=\frac{\Gamma(\lambda)\,e^{-i\pi(2\lambda+1)/4}}{\sqrt{2\pi\omega}}\left(\frac{2}{\omega}\right)^{\lambda}\nonumber\\
&\quad\times\sum_{l=0}^{\infty}(l+\lambda)\left(S_{l}-1\right)C_{l}^{\lambda}(\cos\theta),\nonumber\\
\dsdo&=|h(\theta)|^{2}.
\end{align}
Because of the long range of the gravitational field, the series (\ref{eq:dsdo}) converges poorly. With the phase shifts of the high multipoles, $S_{l}-1\simeq2i\delta_{l}\propto\nu^{4-D}$, the summand behaves as $\nu^{(5-D)/2}$ times an oscillating factor at a fixed angle, so that for $D=5$ the series does not converge in the ordinary sense and for $D=6$, it converges only conditionally, and in the backward direction, where $C_{l}^{\lambda}(-1)=O(\nu^{D-4})$, the summand grows as $(-1)^{l}\nu$. The sum is defined by the Abel summation, which corresponds physically to a wave packet of finite width. In four dimensions, this problem is usually handled by the series reduction of Yennie, Ravenhall, and Wilson~\cite{Yennie1954}. Here, we proceed differently, since the large-$l$ behavior of the phase shifts is known analytically in $D\geq5$. We subtract from the coefficients their large-$\nu$ expansion, which is obtained from the Born approximation for each term of the metric function, and we add it back in closed form using the generating function of the Gegenbauer polynomials. The expansion and the summation are described in Appendix~\ref{app:series}.

The forward divergence of the wave cross section follows from this construction. At small angles, the sum is dominated by the large multipoles, and the leading term of the expansion gives
\begin{equation}\label{eq:forward}
\dsdo\simeq \frac{C_{D}\,M^{2}\omega^{D-4}}{\theta^{4}},\qquad C_{D}=\frac{2^{D-2}\,\Gamma^{2}(D/2)}{\Gamma^{2}(D-2)},
\end{equation}
for $\theta\to0$ and $D\geq5$, with $C_{5}=9\pi/8$. The forward divergence is $\theta^{-4}$ in any dimension, its coefficient does not depend on the charge, and it grows with the frequency as $\omega^{D-4}$. This differs from the classical law (\ref{eq:small}), whose exponent is steeper and whose coefficient does not depend on $\omega$. The two laws describe different orders of limits. The classical law is the eikonal limit, which requires the phase of the partial waves at the relevant multipole, $\nu=\omega b(\theta)$, to be large, and with the universal phase shifts (\ref{eq:deltaweak}) this condition reads $\theta\gg\theta_{*}$ with $\theta_{*}\propto\omega^{-(D-3)/(D-4)}$. For $\theta\ll\theta_{*}$, the wave law (\ref{eq:forward}) holds. For $D=4$, the phase shifts grow logarithmically, both laws are $\theta^{-4}$, and no such transition exists.

\subsection{Semiclassical approximations}

In the high-energy regime (for large $\omega$), the wavelength is almost negligible relative to the horizon scale of the BH. Therefore, in this regime massless scalar waves propagate along the null geodesics \cite{Decanini:2009mu}. Therefore, one may use the classical capture cross section of the massless particle by the BHs as the geometric cross section of the scalar rays. For large $\omega$, the classical deflection angle is related to the phase shift by $\Theta(b)=-2\,d\delta/d\nu$ at $\nu=\omega b$, where the sign reflects our convention that the deflection angle of an attractive field is positive. In the weak field, Eq.~(\ref{eq:weak}) gives the universal leading behavior \cite{Marinho:2016ixt}
\begin{equation}\label{eq:deltaweak}
\delta(\nu)\simeq\frac{\sqrt{\pi}\,\Gamma(\tfrac{D}{2})}{2(D-4)\Gamma(\tfrac{D-1}{2})}\frac{M\omega^{D-3}}{\nu^{D-4}}\qquad(\nu\gg\omega\bc),
\end{equation}
which depends on the mass only, so that the forward scattering is universal. The corrections to Eq.~(\ref{eq:deltaweak}), which contain the charge, are given in Appendix~\ref{app:series}. The strong-deflection regime is related to the Regge poles of the S-matrix, $\nu_{n}\simeq\omega\bc+i\eta_{c}(n+1/2)$, which are related to the quasinormal frequencies of the same BH and have been used to describe the absorption cross section and the quasinormal modes of higher-dimensional BHs~\cite{Decanini2011a,Decanini2011b}. We do not use this approach here.

The oscillations of the cross section near the backward direction are described by the glory approximation~\cite{Matzner1985}. Using the asymptotics of the Gegenbauer polynomials for large $l$~\cite{Szego1975} and the stationary-phase method, we generalize it to $D$ dimensions (Appendix~\ref{app:glory}),
\begin{equation}\label{eq:gloryD}
\dsdo\simeq2\pi\omega\bg^{D-2}\left|\frac{db}{d\theta}\right|_{\theta=\pi}\frac{J^{2}_{(D-4)/2}(\omega\bg\sin\theta)}{\sin^{D-4}\theta},
\end{equation}
where $J_{\mu}$ is the Bessel function. For $D=4$, this is the well-known formula with $J_{0}$~\cite{Matzner1985}. The expression (\ref{eq:gloryD}) contains two geometric quantities: $\bg$ which fixes the positions of the fringes, and $|db/d\theta|_{\theta=\pi}$ which fixes their intensity, and its average over the oscillations reproduces the two-branch classical cross section (\ref{eq:classical}) near $\theta=\pi$. The scattering by higher-dimensional backgrounds has also been considered in Ref.~\cite{Anacleto2023}, where the field is expanded in the ordinary spherical harmonics and the amplitude in the Legendre polynomials, and in that expansion, the four-dimensional formula with $J_{0}$ is the appropriate one. For the planar wave expanded on $S^{D-2}$, the Gegenbauer expansion leads to Eq.~(\ref{eq:gloryD}). Expanding the Bessel function at small argument, we obtain the backward cross section,
\begin{equation}\label{eq:back}
\dsdo\Big|_{\theta=\pi}=\frac{2\pi\,\omega^{D-3}\bg^{2D-6}}{2^{D-4}\,\Gamma^{2}\!\left(\tfrac{D-2}{2}\right)}\left|\frac{db}{d\theta}\right|_{\theta=\pi},
\end{equation}
which is finite in any dimension and grows as $\omega^{D-3}$. The zeros of Eq.~(\ref{eq:gloryD}) are the zeros of $J_{\mu}$ with $\mu=(D-4)/2$, so that the first minimum exists only if $\omega\bg$ exceeds the first zero $j_{\mu,1}$, which grows with the dimension, and in the region $1/(\omega\bg)\ll\pi-\theta\ll1$, the fringes are equally spaced with the period
\begin{equation}\label{eq:dtheta}
\Delta\theta\simeq\frac{\pi}{\omega\bg}\,,
\end{equation}
so that the fringes are wider for smaller $\bg$ and smaller $\omega$.

We solve Eq.~(\ref{eq:RW}) numerically from a point close to the horizon with the ingoing boundary condition, up to a large radius, where the solution is matched with the Riccati-Hankel functions, including the first-order phase correction of the tail of the potential beyond the matching radius. The multipoles up to $l_{\max}=45\,\omega+10$ are computed, and the series is summed as described in Appendix~\ref{app:series}. The flux conservation, the stability of the S-matrix elements with respect to the matching radius, the integration tolerance, and the starting point near the horizon (including the extremal case, where the horizon is double), and the stability of the summed cross sections with respect to the number of multipoles and to the way the tail is treated have been checked, and the backward cross sections, which are the most sensitive ones, are determined within $0.1\%$. 

\section{Results and discussions}\label{sec:results}

\subsection{Phase shifts}

\begin{figure*}[!htb]
\includegraphics[width=0.48\linewidth]{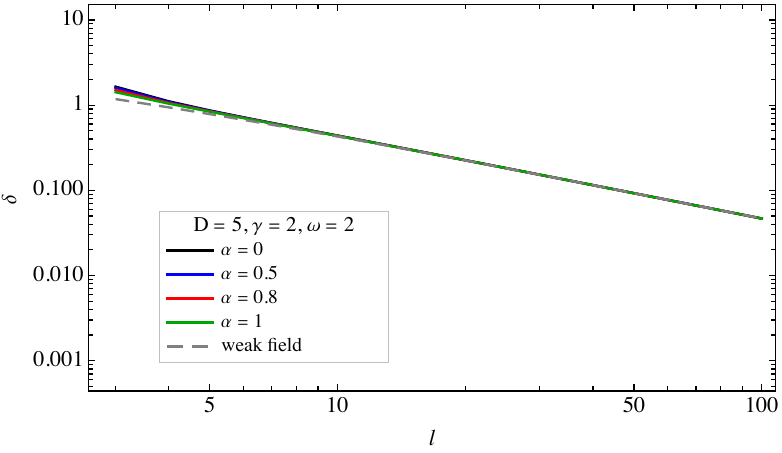}
\includegraphics[width=0.48\linewidth]{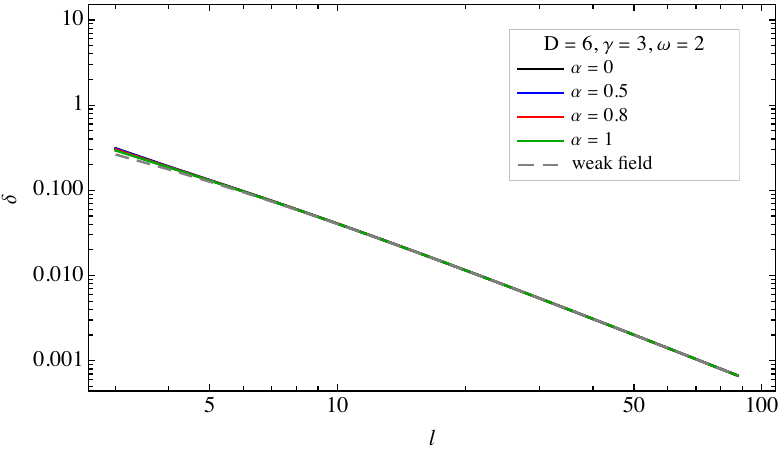}
\caption{Real parts of the phase shifts for $\omega=2$ and different $\alpha$ for $D=5$, $\gamma=2$ and $D=6$, $\gamma=3$ (left and right panels). The dashed gray line is the weak-field law (\ref{eq:deltaweak}).}
\label{fig:delta}
\end{figure*}

The real parts of the phase shifts for $\omega=2$ are shown in Fig.~\ref{fig:delta}. At large $l$, the phase shifts of all BHs approach the universal law (\ref{eq:deltaweak}), since the waves with large $l$ probe only the long-range part of the potential, which is determined by the mass. The approach is slow, however, and the charge enters the corrections to the universal law, which are given in Appendix~\ref{app:series}. The charge affects mainly the low multipoles, $\nu\lesssim\omega\bc$, which are responsible for the scattering at large angles and for the absorption. For $D=5$, the phase shifts decay as $1/\nu$, so that the terms of the series (\ref{eq:dsdo}) do not decay, while for $D=6$, they decay as $1/\nu^{2}$.

\subsection{Scattering cross section}

\begin{figure*}[!htb]
\includegraphics[width=0.48\linewidth]{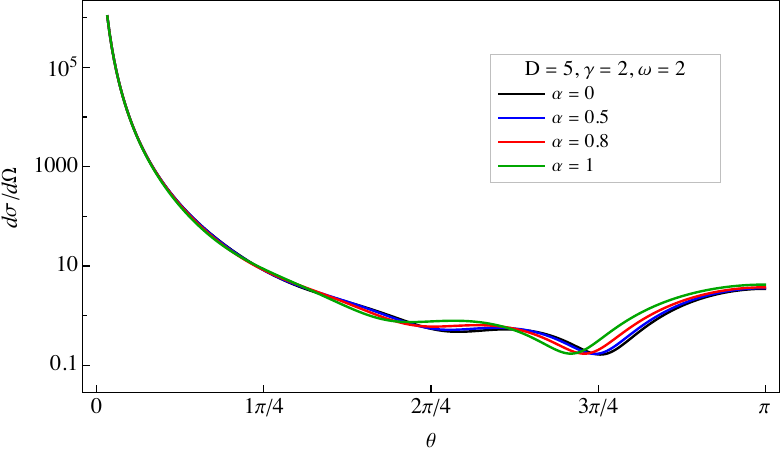}
\includegraphics[width=0.48\linewidth]{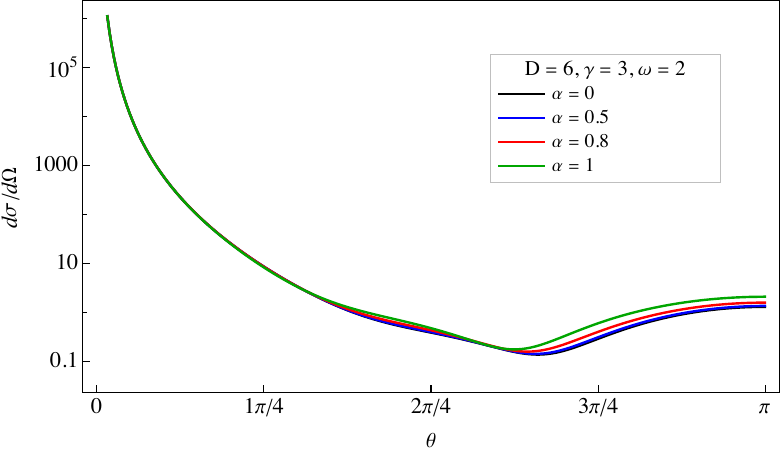}
\caption{Scattering cross section of the scalar field with $\omega=2$ for $D=5$, $\gamma=2$ and $D=6$, $\gamma=3$ (left and right panels).}
\label{fig:scat}
\end{figure*}

Fig.~\ref{fig:scat} shows the scattering cross section for $\omega=2$ and several values of $\alpha$. In all cases, the cross section diverges in the forward direction, decreases at intermediate angles, and shows interference fringes toward the backward direction. The forward part of the cross section does not depend on $\alpha$, since it is determined by the large-$l$ phase shifts, while the fringes become wider with increasing charge, since $\bg$ decreases with $\alpha$ [Eq.~(\ref{eq:dtheta})], as in four dimensions~\cite{Macedo2015}. In five dimensions, the cross section is finite and nonzero at $\theta=\pi$, in agreement with Eq.~(\ref{eq:back}), while in six dimensions, the glory pattern contains fewer oscillations at the same $\omega$, because $\bg$ is smaller in higher dimensions, and the fringes are wider.

The patterns of Fig.~\ref{fig:scat} contain fewer fringes than the four-dimensional ones at comparable $\omega M$~\cite{Macedo2014,Macedo2015,Paula2020} for two reasons. First, the number of fringes is set by $\omega\bg$ [Eq.~(\ref{eq:dtheta})], and in four dimensions $\bg\simeq5.2M$, so that $\omega M=3$ corresponds to $\omega\bg\simeq16$, while in our units $\omega=2$ corresponds to $\omega\bg\simeq4$, i.e., to $\omega M\simeq0.7$ in four dimensions, where only one or two fringes appear as well. Second, the oscillations at intermediate angles come from the interference of the direct ray with the rays that pass on the other side of the BH, whose amplitudes are suppressed by $e^{-\pi\eta_{c}}$ (Sec.~\ref{sec:geo}). Since $\eta_{c}=1$ for the Schwarzschild BH and $\eta_{c}=\sqrt{D-3}$ for the Tangherlini BH, these rays are several times weaker in $D=5,6$, and the forward region is the smooth universal tail (\ref{eq:forward}) rather than the logarithmic Coulomb-like interference of four dimensions. The smoother patterns are therefore a real feature of the higher-dimensional BHs.

In the forward direction, the cross sections follow the wave law (\ref{eq:forward}), whose coefficient we have checked against the numerical results. The classical exponent (\ref{eq:small}) is approached only at intermediate angles and at the higher frequencies. For $\omega=1$ in five dimensions, where $\omega\bc\simeq2$, the local slope of the cross section stays at four up to the region of the glory fringes, for $\omega=2$ and $3$ it approaches the classical value in an intermediate range of angles before the fringes set in, and in six and seven dimensions at $\omega=2$, it rises above four but does not reach the classical values. Since $\omega\bc$ is only of order a few at these frequencies, we regard this behavior as the onset of the transition toward the classical regime, which requires higher frequencies in higher dimensions.

\begin{figure}[!htb]
\includegraphics[width=\columnwidth]{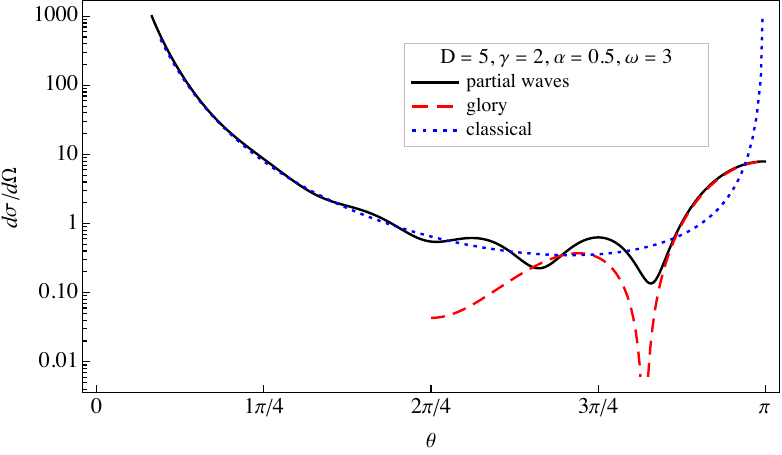}
\caption{Scattering cross section for $D=5$, $\gamma=2$, $\alpha=0.5$, and $\omega=3$, namely, the partial-wave result (solid), the glory approximation (dashed), and the classical cross section (dotted).}
\label{fig:glory}
\end{figure}

In Fig.~\ref{fig:glory}, the partial-wave result is compared with the glory approximation (\ref{eq:gloryD}) and the classical cross section (\ref{eq:classical}). The glory approximation reproduces the intensity near the backward direction and the position of the last minimum, while the classical cross section describes the averaged flux at small and intermediate angles, where the interference is weak, and the envelope of the fringes near the backward direction. For the weakly charged BHs, the backward cross section (\ref{eq:back}) agrees with the partial waves within a few percent for $\omega\bg\gtrsim3$ in $D=5,6,7$ and within about one percent for $\omega\bg\gtrsim5$ in the cases we have tested, while for the extremal five-dimensional BH the glory formula overestimates the backward intensity by about five percent, which is the size of the interference with the first winding ray discussed in Sec.~\ref{sec:geo}. The secondary maxima are not reproduced by the single-ray formula, because at these angles the glory ray interferes with the other branches of the deflection function. The comparison for several frequencies and charges is available upon a reasonable request.

\subsection{Dependence on $\gamma$, $\omega$, and $D$}

\begin{figure}[!htb]
\includegraphics[width=\columnwidth]{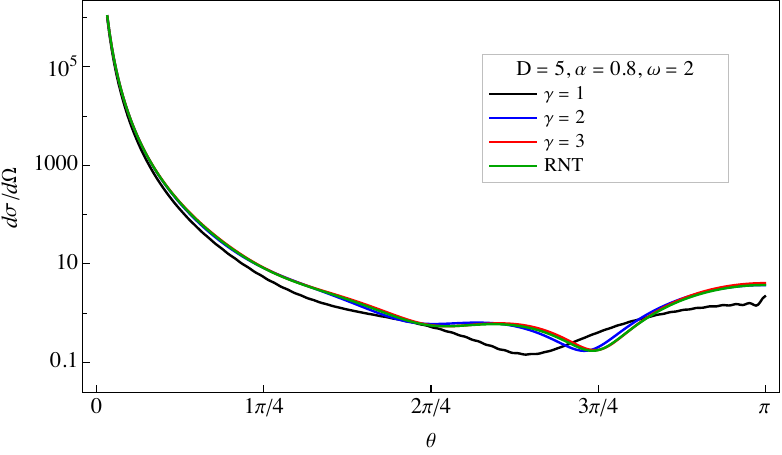}
\caption{Scattering cross section for $D=5$, $\alpha=0.8$, and $\omega=2$ for $\gamma=1$ (black), $\gamma=2$ (blue), $\gamma=3$ (red), and the RNT BH.}
\label{fig:gamma}
\end{figure}

The role of the parameter $\gamma$ is shown in Fig.~\ref{fig:gamma} for the five-dimensional BH with $\alpha=0.8$. The shape of the cross section is qualitatively the same for all $\gamma$, and the differences are concentrated in the positions and depths of the last minima, which follow the differences in $\bg$ in Fig.~\ref{fig:bc}. The model with $\gamma=1$, whose charge correction decays slowest, deviates most from the others, while the Bardeen-like ($\gamma=2$) and Hayward-like ($\gamma=3$) models are close to the RNT BH. This confirms that the geometric quantities $\bg$ and $|db/d\theta|_{\theta=\pi}$, rather than the details of the metric, control the backward scattering.

\begin{figure}[!htb]
\includegraphics[width=\columnwidth]{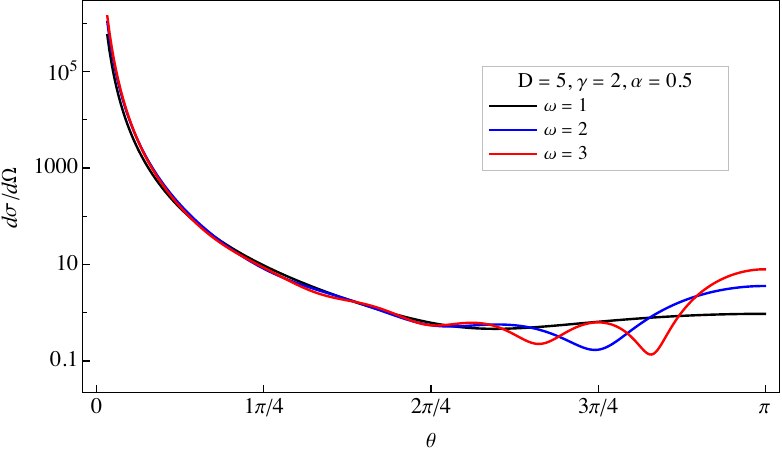}
\caption{Scattering cross section for $D=5$, $\gamma=2$, and $\alpha=0.5$ for $\omega=1$ (black), $\omega=2$ (blue), and $\omega=3$ (red).}
\label{fig:omega}
\end{figure}

The frequency dependence is shown in Fig.~\ref{fig:omega}. With increasing $\omega$, the wavelength becomes smaller than the size of the BH, the number of fringes grows according to Eq.~(\ref{eq:dtheta}), and the backward cross section increases, as predicted by Eq.~(\ref{eq:back}). The forward cross section also grows, as $\omega^{D-4}$ according to Eq.~(\ref{eq:forward}), i.e., linearly with $\omega$ in five dimensions, which is hardly visible on the logarithmic scale of the figure. In the limit $\omega\to\infty$, the cross section is expected to approach the classical one averaged over the fringes, outside a forward cone that shrinks with the frequency and outside a backward region of angular size $1/(\omega\bg)$, where the classical cross section is singular and the wave cross section stays finite. For low frequencies, $\omega\bc\lesssim1$, the interference is washed out, and the cross section is smooth.

\begin{figure}[!htb]
\includegraphics[width=\columnwidth]{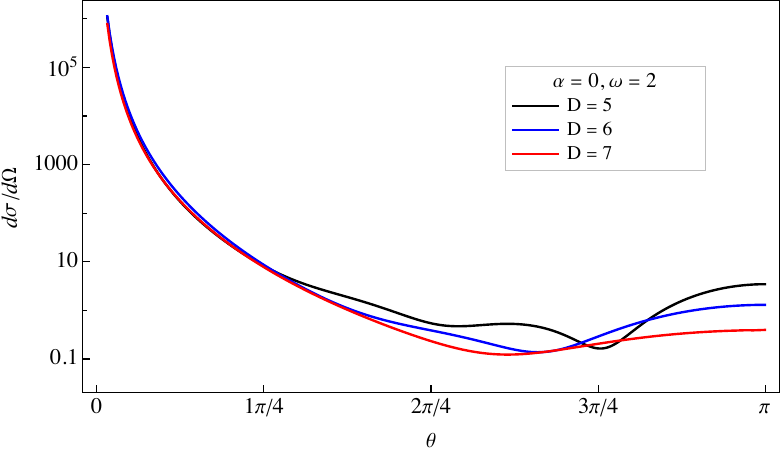}
\caption{Scattering cross section for the Schwarzschild-Tangherlini BH ($\alpha=0$) and $\omega=2$ in $D=5$ (black), $D=6$ (blue), and $D=7$ (red), in units of $M^{(D-2)/(D-3)}$.}
\label{fig:dim}
\end{figure}

Finally, the effect of the number of dimensions is shown in Fig.~\ref{fig:dim} for the Schwarzschild-Tangherlini BH. Since the unit of length depends on $D$, we compare BHs with the same mass parameter, i.e., $\omega=2$ in units of $M^{-1/(D-3)}$. In the forward direction, all curves follow the $\theta^{-4}$ law with the coefficients $C_{D}\omega^{D-4}$. Since $\bc$ decreases with $D$ in these units, $\omega\bg$ decreases, and the first zero of the Bessel function moves to larger values with $D$, so that the glory pattern contains fewer fringes in higher dimensions, and for $D=6$ and $D=7$ at this frequency, the glory formula has no zero at all, in agreement with the smooth bumps of Fig.~\ref{fig:dim}. The backward cross section also decreases with $D$. This decrease is not caused by the factor $(\omega\bg/2)^{D-4}$ of the Bessel function expansion, which grows with $D$, but by the Jacobian $|db/d\theta|_{\theta=\pi}$, which decreases by an order of magnitude from $D=5$ to $D=7$. The circular null geodesic is more unstable in higher dimensions ($\eta_{c}=\sqrt{D-3}$), so that the glory ray is closer to the critical one, and the bundle of rays scattered backwards is thinner, $|db/d\theta|\simeq\eta_{c}\bc(\bg/\bc-1)\propto e^{-\pi\eta_{c}}$. The fringes can be resolved in higher dimensions only at higher frequencies.

\subsection{Regular versus singular black holes}

Finally, let us discuss whether the exterior geometry of the regular BH can be distinguished from that of the RNT BH by the scalar waves. Since the fringes are controlled by $\bg$, two BHs with the same $\bg$ should have similar scattering patterns~\cite{Macedo2016,Paula2026,Tang2026}. For the five-dimensional Bardeen-like BH with $\alpha=0.7$, the RNT BH with the same glory impact parameter has $\alpha\simeq0.94$. The two BHs have almost the same critical impact parameter and capture cross section, but the glory Jacobians $|db/d\theta|_{\theta=\pi}$ differ by about six percent and the horizon areas by about fourteen percent. Matching $\bg$ therefore matches the positions of the fringes, but neither their intensity, which is controlled by the Jacobian, nor the absorption at low frequencies, which is controlled by the horizon area.

\begin{figure*}[!htb]
\includegraphics[width=0.48\linewidth]{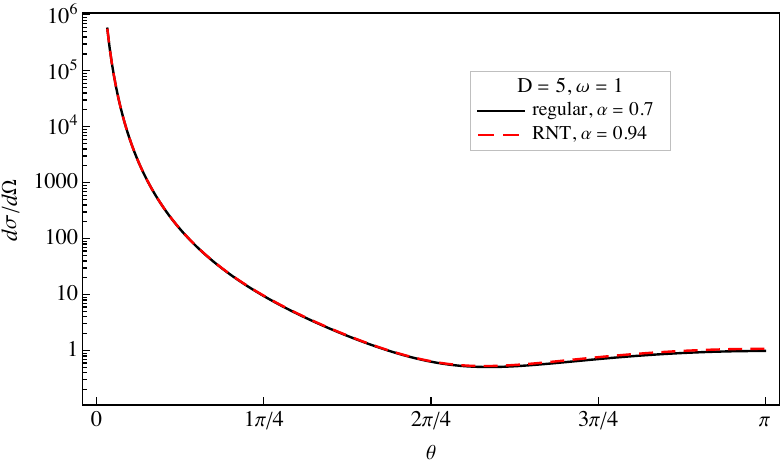}
\includegraphics[width=0.48\linewidth]{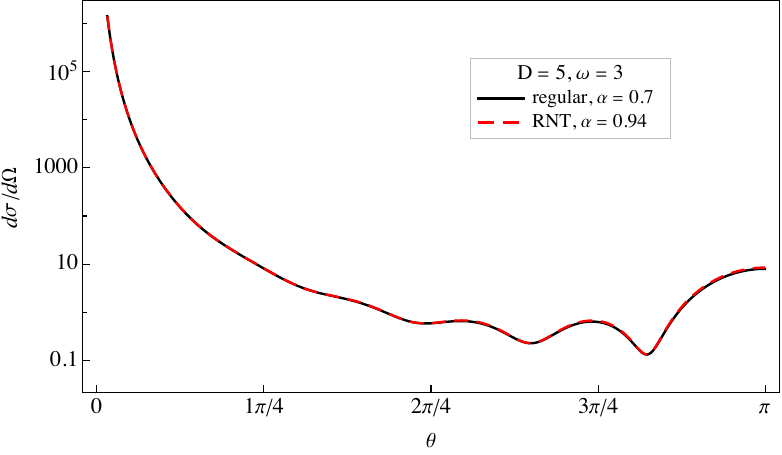}
\caption{Scattering cross sections of the five-dimensional Bardeen-like BH with $\alpha=0.7$ (solid) and of the RNT BH with the same $\bg$ (dashed) for $\omega=1$ and $\omega=3$ (left and right panels).}
\label{fig:rnt}
\end{figure*}

Fig.~\ref{fig:rnt} shows that the two scattering cross sections are close at all angles for both frequencies. Their relative difference vanishes in the forward direction, where the cross section is universal, is at the level of a few percent at intermediate angles, and is largest near the backward direction, where it reaches about nine percent at $\omega=1$ and about six percent at $\omega=3$, the latter being the difference of the Jacobians, in agreement with the glory formula (\ref{eq:back}). Thus, the two exterior geometries are approximately degenerate for the scalar waves, at the level of ten percent at the angles and frequencies considered, but not exactly degenerate. The reason for the approximate degeneracy is that the cross section depends on the BH mainly through the long-range tail, which is universal, and through the glory ray. We stress that the scattering problem with the ingoing boundary condition at the horizon compares two exterior geometries and does not probe the region behind the horizon, so that the regularity of the core itself is not tested by it. Because $\bg$ of the Maxwellian regular BH is smaller than that of the RNT BH at the same $\alpha$, the regular BH can be mimicked only by an RNT BH with a larger charge, and within the RNT family at the same mass parameter, the near-extremal regular BHs have no partner with the same $\bg$.

\begin{figure}[!htb]
\includegraphics[width=\columnwidth]{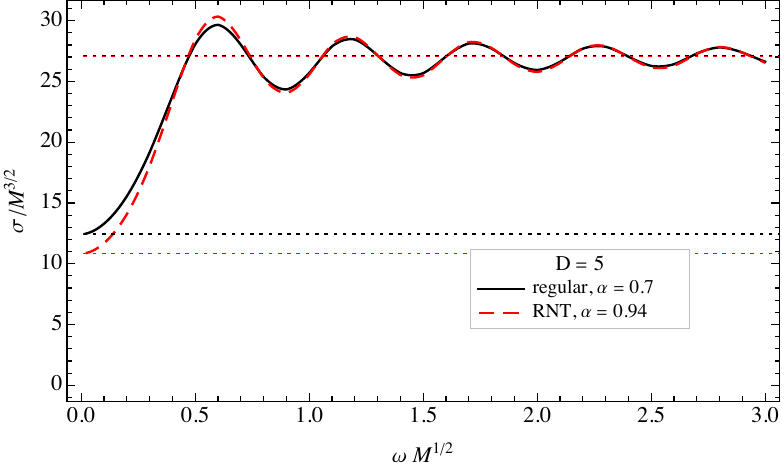}
\caption{Absorption cross sections of the five-dimensional Bardeen-like BH with $\alpha=0.7$ (solid) and of the RNT BH with the same $\bg$ (dashed). The upper dotted lines are the capture cross sections, and the lower dotted lines are the horizon areas.}
\label{fig:rntabs}
\end{figure}

The absorption cross section is obtained from the same S-matrix elements,
\begin{equation}\label{eq:abs}
\sigma_{\rm abs}=\frac{(2\pi)^{D-2}}{\Omega_{D-2}\,\omega^{D-2}}\sum_{l}N_{l}\left(1-|S_{l}|^{2}\right),
\end{equation}
where 
\begin{eqnarray}
    N_{l}=\frac{(2l+D-3)(l+D-4)!}{(D-3)!\,l!}
\end{eqnarray}
is the multiplicity of the harmonics. Fig.~\ref{fig:rntabs} shows the absorption cross sections of the two BHs. At high frequencies, they oscillate around the capture cross sections, which are almost equal for the two BHs, and their relative difference is below one percent for $\omega\geq1$. At low frequencies, the absorption cross section of a massless scalar field tends to the horizon area~\cite{Das1997}, and our results approach this limit for both BHs. Since the horizon areas differ by about fourteen percent, the low-frequency absorption distinguishes the two BHs, while the high-frequency absorption does not. The glory Jacobian and the horizon area are thus two distinct probes of the exterior geometry, the former being determined by the deflection function near $\bg$ and the latter by the horizon radius.

\section{Conclusion}\label{sec:conclusion}

In this paper, we have studied the scattering of massless scalar waves by the higher-dimensional regular BHs in general relativity coupled to the NED~\cite{Toshmatov2026HD,Toshmatov:2026xtc}, which unify the Bardeen and Hayward models. The scalar field is a neutral test field in the bulk, whose rays in the geometric-optics limit are the null geodesics of the spacetime, so that the unstable circular null geodesics of the background geometry, and not the photonsphere of the effective geometry, govern the scattering. We have combined the classical scattering of null geodesics, the glory approximation, and the numerical partial-wave analysis.

The charge parameter decreases the critical and the glory impact parameters, the parameter $\eta_{c}$ of the circular null geodesic, and the deflection of the null geodesics. At the same normalized charge, the Maxwellian regular BHs have smaller $\bc$ and $\bg$ than the RNT BH, but at the same charge $q$, they are slightly larger, as in four dimensions. These quantities decrease with the dimension and with decreasing $\gamma$.

We have derived the glory formula in $D$ dimensions for the planar wave expanded on $S^{D-2}$ and tested it against the partial waves. It reproduces the backward cross section of the weakly charged BHs within a few percent and the position of the first fringe at the higher frequencies, while for the near-extremal BHs its accuracy is limited by the winding rays. The backward cross section is finite in any dimension and grows as $\omega^{D-3}$. The charge widens the fringes, and in higher dimensions, the pattern at a fixed frequency has fewer and wider fringes and a weaker backward intensity, because the Jacobian of the glory decreases with $\eta_{c}$. At the same $\omega\bg$, the number of fringes is the same as in four dimensions, but the oscillations at intermediate angles are weaker, since the rays that pass on the other side of the BH are suppressed by $e^{-\pi\eta_{c}}$.

At a fixed frequency, the forward divergence of the wave cross section is $\theta^{-4}$ in any dimension, with a coefficient that is fixed by the mass and grows as $\omega^{D-4}$. The classical law $\theta^{-(D-2)^{2}/(D-3)}$ is the eikonal limit, and it is approached only at intermediate angles and at the higher frequencies, which are higher in higher dimensions. The phase shifts of the high multipoles approach a universal law fixed by the mass, which explains why the forward scattering does not depend on the charge, and the partial-wave series in $D\geq5$ can be summed by subtracting the known large-$l$ behavior of its coefficients.

A five-dimensional regular BH and an RNT BH of a different charge with the same glory impact parameter have scattering cross sections that differ by less than ten percent at all angles, with the largest difference in the backward direction, and absorption cross sections that differ by less than one percent at high frequencies and by the ratio of the horizon areas at low frequencies. Thus, the two exterior geometries are approximately, but not exactly, degenerate for the scalar waves, and the degeneracy is lifted by the backward intensity and by the low-frequency absorption. These conclusions are the $D$-dimensional counterparts of those obtained in four dimensions~\cite{Macedo2016,Paula2026,Tang2026}.

It would be interesting to extend this work to the gravitational field, to massive and charged scalar fields~\cite{Dolan2024}, and to rotating higher-dimensional regular BHs, for which the glory is expected to be modified by the frame dragging. The EM waves require a separate treatment, since in the NED, the light rays and the EM perturbations propagate along the null geodesics of the effective geometry rather than of the spacetime itself~\cite{Toshmatov2018axial,Toshmatov2018polar,Toshmatov2019relax}, so that the photonsphere, the lensing, and the scattering of the EM waves are governed by the effective geometry. This problem will be considered in a forthcoming paper.

\begin{acknowledgments}
This research was partially supported by the Ministry of Higher Education, Science, and Innovation of the Republic of Uzbekistan under Grants No. FL-9524114991 and No. FL-10425067111.

In preparing this work, the authors used Claude (Fable 5.1) as an assistant for symbolic and numerical cross-checks of the calculations, for drafting parts of the text, and for writing the scripts to produce figures. All derivations, results, and the final text were checked, revised, and approved by the authors, who take full responsibility for the content of the paper.
\end{acknowledgments}

\appendix
\section{Derivation of the glory formula in $D$ dimensions}\label{app:glory}

We start from the amplitude (\ref{eq:dsdo}) and consider the angles close to the backward direction, $\theta=\pi-\epsilon$ with $\epsilon\ll1$. We use $C_{l}^{\lambda}(-x)=(-1)^{l}C_{l}^{\lambda}(x)$ and the Hilb-type asymptotic formula for the Gegenbauer polynomials at large $\nu=l+\lambda$~\cite{Szego1975},
\begin{equation}\label{eq:hilb}
C_{l}^{\lambda}(\cos\epsilon)\simeq\frac{\sqrt\pi\,2^{1/2-\lambda}}{\Gamma(\lambda)}\,\nu^{\lambda-1/2}\,\frac{(\epsilon/\sin\epsilon)^{1/2}}{(\sin\epsilon)^{\lambda-1/2}}\,J_{\lambda-1/2}(\nu\epsilon),
\end{equation}
which reduces to the familiar $P_{l}(\cos\epsilon)\simeq J_{0}(\nu\epsilon)$ for $\lambda=1/2$. In the semiclassical regime, we write $S_{l}=e^{2i\delta(\nu)}$ and $(-1)^{l}=e^{i\pi(\nu-\lambda)}$, replace the sum by an integral over $\nu$, and evaluate it by the stationary-phase method. The phase $\Phi(\nu)=2\delta(\nu)+\pi\nu$ is stationary where $2\delta'(\nu)=-\pi$, i.e., where $\Theta(b)=\pi$, so that the stationary point is the glory ray, $\nu_{g}=\omega\bg$, and both branches $\Theta=\theta$ and $\Theta=2\pi-\theta$ meet there. The term $-1$ in $S_{l}-1$ has no stationary point and contributes only to the forward direction. Since the Bessel function varies slowly in comparison with $e^{i\Phi}$ near $\nu_{g}$, it is evaluated at $\nu_{g}$, and the Gaussian integral gives
\begin{equation}\label{eq:statphase}
\left|\int d\nu\,e^{i\Phi(\nu)}\right|\simeq\sqrt{\frac{2\pi}{|\Phi''(\nu_{g})|}}=\sqrt{2\pi\omega\left|\frac{db}{d\theta}\right|_{\theta=\pi}},
\end{equation}
where we have used $\Phi''=2\delta''=-\omega^{-1}d\Theta/db$. Collecting the factors, with $|K|^{2}=\Gamma^{2}(\lambda)(2/\omega)^{2\lambda}/(2\pi\omega)$ for the prefactor of Eq.~(\ref{eq:dsdo}) and $\nu_{g}^{2\lambda+1}/\omega^{2\lambda}=\omega\bg^{D-2}$, we obtain
\begin{equation}\label{eq:glory_app}
|h|^{2}\simeq2\pi\omega\bg^{D-2}\left|\frac{db}{d\theta}\right|_{\theta=\pi}\frac{J^{2}_{\lambda-1/2}(\omega\bg\epsilon)}{(\sin\epsilon)^{2\lambda-1}}\,,
\end{equation}
which is Eq.~(\ref{eq:gloryD}) once $\epsilon$ in the argument of the Bessel function is replaced by $\sin\theta$, as in four dimensions~\cite{Matzner1985}. The winding rays, $\Theta=(2k+1)\pi$, give additional stationary points with the weights $|db/d\theta|_{k}\propto e^{-(2k+1)\pi\eta_{c}}$, which are neglected.

\section{Large-$l$ expansion and summation of the partial-wave series}\label{app:series}

The first Born approximation gives the phase shift of each term $-\mu/r^{p}$ of the metric function exactly to first order in $\mu$. In the tortoise coordinate normalized so that $r_{*}\to r$ at infinity, the potential (\ref{eq:V}) differs from the free one by
\begin{equation}\label{eq:dV}
\Delta V=-\frac{\mu}{r_{*}^{p+2}}\left[\frac{p+1}{p-1}\left(\nu^{2}-\lambda^{2}\right)+\frac{(D-2)p(D-3-p)}{2(p-1)}\right],
\end{equation}
where the second term vanishes for the mass term, $p=D-3$, and the Born phase shift, $\delta^{(1)}=-\omega^{-1}\int_{0}^{\infty}\Delta V\,\hat{\jmath}_{\nu}^{2}(\omega r_{*})dr_{*}$ with the Riccati-Bessel function $\hat{\jmath}_{\nu}$, follows from the integral of the product of two Bessel functions~\cite{DLMF},
\begin{align}\label{eq:born}
\delta^{(1)}(\nu)&=\frac{\pi\mu\omega^{p}}{2}\left[\frac{p+1}{p-1}\left(\nu^{2}-\lambda^{2}\right)+\frac{(D-2)p(D-3-p)}{2(p-1)}\right]\nonumber\\
&\quad\times\frac{\Gamma(p+1)\,\Gamma(\nu-\tfrac{p}{2})}{2^{p+1}\,\Gamma^{2}(\tfrac{p+2}{2})\,\Gamma(\nu+\tfrac{p}{2}+1)},
\end{align}
which is a rational function of $\nu$ for integer $p$. Its leading term coincides with the integral of Eq.~(\ref{eq:weak}), $\delta=(\omega/2)\int_{\nu/\omega}^{\infty}\Theta\,db$, and the ratio of the Gamma functions provides the corrections. The terms of second order in the metric function are taken from the second-order expansion of the deflection angle, which for the two Maxwellian models reads
\begin{align}
\Theta&=\frac{3\pi M}{4b^{2}}+\frac{\pi M}{b^{4}}\left(\frac{105M}{64}-\frac{15Q^{2}}{8}\right)+O(b^{-6}),\quad D=5,\nonumber\\
\Theta&=\frac{8M}{3b^{3}}+\frac{\pi M}{b^{6}}\left(\frac{315M}{128}-\frac{175Q^{3}}{96}\right)+O(b^{-9}),\quad D=6.\label{eq:weak2}
\end{align}
The resulting expansions of the phase shifts are
\begin{align}
\delta&=\frac{3\pi M\omega^{2}}{8\nu}-\frac{5\pi MQ^{2}\omega^{4}}{16}\frac{\nu^{2}-17/5}{\nu(\nu^{2}-1)(\nu^{2}-4)}\nonumber\\
&\quad+\frac{35\pi M^{2}\omega^{4}}{128\nu^{3}}+O(\nu^{-5}),\qquad D=5,\ \gamma=2,\nonumber\\
\delta&=\frac{2M\omega^{3}}{3(\nu^{2}-1/4)}+\frac{\pi M\omega^{6}}{10\nu^{5}}\left(\frac{315M}{128}-\frac{175Q^{3}}{96}\right)\nonumber\\
&\quad+O(\nu^{-6}),\qquad D=6,\ \gamma=3,\nonumber\\
\delta&=\frac{5\pi M\omega^{4}}{32\nu(\nu^{2}-1)}+O(\nu^{-7}),\qquad D=7,\ \alpha=0,\label{eq:deltaasym}
\end{align}
and the expansions for the RNT metric and for $\gamma=1$ are built in the same way. In five dimensions, the mass term has no correction, and the first correction to the universal law contains the charge, while in six dimensions, the first correction is the mass term $M\omega^{3}/(6\nu^{4})$, and the charge enters only at the next order.

To sum the series (\ref{eq:dsdo}), we write the coefficients as $c_{l}=\nu(S_{l}-1)$ and subtract their large-$\nu$ expansion $T(\nu)=\sum_{k=0}^{D-2}c_{k}\nu^{-k}$, obtained by expanding $\nu(e^{2i\delta}-1)$ with the phase shifts (\ref{eq:deltaasym}),
\begin{equation}\label{eq:split}
h(\theta)\simeq K\left[\sum_{l=0}^{l_{c}}\left(c_{l}-T(\nu_{l})\right)C_{l}^{\lambda}(x)+\sum_{l=0}^{\infty}T(\nu_{l})C_{l}^{\lambda}(x)\right],
\end{equation}
where $x=\cos\theta$ and $K$ is the prefactor of Eq.~(\ref{eq:dsdo}). The second sum is a combination of $\Sigma_{k}(x)=\sum_{l}C_{l}^{\lambda}(x)/\nu_{l}^{k}$, which follow from the generating function $\sum_{l}C_{l}^{\lambda}(x)t^{l}=(1-2xt+t^{2})^{-\lambda}$ as $\Sigma_{0}(x)=[2(1-x)]^{-\lambda}$ (Abel sum) and, for $k\geq1$,
\begin{equation}\label{eq:Sigmak}
\Sigma_{k}(x)=\frac{1}{(k-1)!}\int_{0}^{1}\frac{u^{\lambda-1}(-\ln u)^{k-1}}{(1-2xu+u^{2})^{\lambda}}\,du,
\end{equation}
which are valid for all $-1\leq x<1$, including the backward direction. The remainder $c_{l}-T(\nu_{l})$ decays as $\nu^{-(D-1)}$, so that the neglected terms of the first sum decay as $\nu^{-3}$ even at $\theta=\pi$. The computed multipoles are used up to the $l_{c}$ at which the remainder reaches the numerical noise of $S_{l}$, and one more fitted term of $T(\nu)$ is included. The residuals follow the expected power laws, and the resulting cross sections are stable with respect to $l_{c}$, to its parity, and to the fitted term, and agree with an Abel-damped evaluation of the backward amplitude. The forward law (\ref{eq:forward}) follows from the leading term of $T(\nu)$ and $\Sigma_{D-5}(x)\simeq\theta^{-2}/(D-5)!$ at $x\to1$.

\bibliography{refs2}

\end{document}